\documentclass{v18windows}

\usepackage{wrapfig}

\def\be{\begin{equation}}
\def\ee{\end{equation}}
\def\bea{\begin{eqnarray}}
\def\eea{\end{eqnarray}}

\newcommand{\Photo}{}

\begin{document}
\vspace*{4cm}
\title{CONSTRUCTING THE ASTRONOMICAL OBSERVATORY AT TAY NGUYEN UNIVERSITY, VIETNAM}

\author{Tan LE-MINH}

\address{Department of Physics, Faculty of Natural Science and Technology, Tay Nguyen University, \\
  567 Le Duan street, Buon Ma Thuot city, Dak Lak province, Vietnam}
  
\author{Lam TRAN-QUOC}

\address{Department of Physics, Faculty of Natural Science and Technology, Tay Nguyen University, \\
  567 Le Duan street, Buon Ma Thuot city, Dak Lak province, Vietnam}
  
\author{ Quang NGUYEN-LUONG }

\address{120 Bloor Street East, Toronto, ON, M4Y 1B7, Canada, 
and at Nagoya City University}

\maketitle
\abstracts{
We are planning to build an observatory that includes a 40 cm optical telescope at Tay Nguyen University (TNU), Daklak Province, Vietnam. In this paper, we report the initial activities of our project. First, to prepare for the scientific exploitation of the observatory, we organized the observational nights for students at the International Center for Interdisciplinary Science and Education, Qui Nhon in April 2017 and at the Nha Trang Observatory, Nha Trang in November 2018. These activities aim to provide students with the skills of astronomical observations and to create their passion in astronomy study. Second, we are organizing the annual Exploration Program in Astrophysics Research (EPAR) at TNU. The goals of EPAR are to promote the teaching and learning activities at TNU and promoting collaboration between TNU and other institutes. Finally, we are conducting joint research project between TNU and Kyoto astronomical observatory of Kyoto university to prepare for the technical details of the observatory.
}

\section{Why do we need an astronomical observatory in Tay Nguyen University?}

In 2017-2018, the Vietnam National Space Center (VNSC) installed two 50 cm optical telescopes in Hanoi and Nha Trang cities. They are about to begin operations. In addition, the International Center for Interdisciplinary Science and Education (ICISE) in Quy Nhon city and VNSC held a workshop on Astronomy Development in Vietnam - Challenges and Opportunities from 31 July to 2 August, 2016. The purpose of the workshop is ``seek to establish a road-map for the development of astronomy and astrophysics in Vietnam". Following this workshop's goal, Tay Nguyen university (TNU) in Dak Lak province, Vietnam has held the special programs for astronomy and strives to attract students to study astronomy.

Dak Lak province is the largest province in the HighLand region in Vietnam. This province is in one of the mountain region of Vietnam which has an average high of 400 - 800 m above sea level and the peaks up to 1500 - 2000 m. It has a high mountain, Chu Yang Sin, with 2442 m high. Dak Lak is about 200 km far from the sea. The dry season begins from November to April each year with quite dry weather and many clear nights.
The TNU located in Daklak is the largest university in the region. The TNU has about 12800 students, among which 600 students are in Faculty of Natural Science and Technology. There are about 80 students in Physics at Department of Physics. Department of Physics has two lecturers and one technician in Astronomy. One lecture, Nguyen Dang Thanh Nhan, now is pursuing a graduate program in Astrophysics at Bonn, Germany. Astronomy subject is taught for the third-year students. It has three credits with 25 hours for lecture and 3 days for observational practice. However, a night-sky observation is still a big gap because we now just have a small optical telescope with 90 mm in diameter.

\begin{wrapfigure}{r}{0.5\textwidth} 
\vspace{-20pt}
  \begin{center}
    \includegraphics[width=0.4\textwidth]{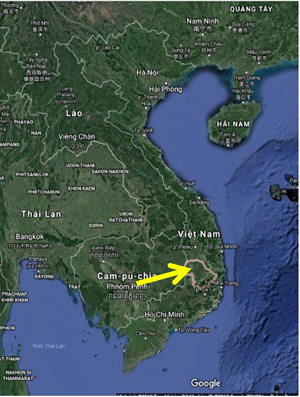}
    \caption{Vietnam map shows the location of DakLak province where TNU resides.}
\label{fig:radish}
  \end{center}
  \vspace{-20pt}
 \vspace{3pt}
\end{wrapfigure}

We plan to start building an astronomical observatory at TNU which will contain a 40 cm optical telescope. The main audience of the observatory is the students to conduct experiments and exercise. In return, the students will conduct public shows to draw the attention of the public to science and to increase the awareness of science. Astronomical observation brings the passion and creativity for students and promote the researching and teaching activities in Astronomy at TNU. The observatory is also used for research purpose such as finding the exoplanets, studying planet formation, observing asteroids, etc. Furthermore, according to the draft of Vietnam High School Physics curriculum in 2017, the Physics program for the 10$^{th}$ grade has a subject of the Earth and Sky with the time of 11 hours. Therefore, the observatory is an ideal destination for High School students in Dak Lak province to practise this subject. The main goals of our project are as follows,

\begin{figure}[htbp]
\centerline{\includegraphics[width=0.95\linewidth]{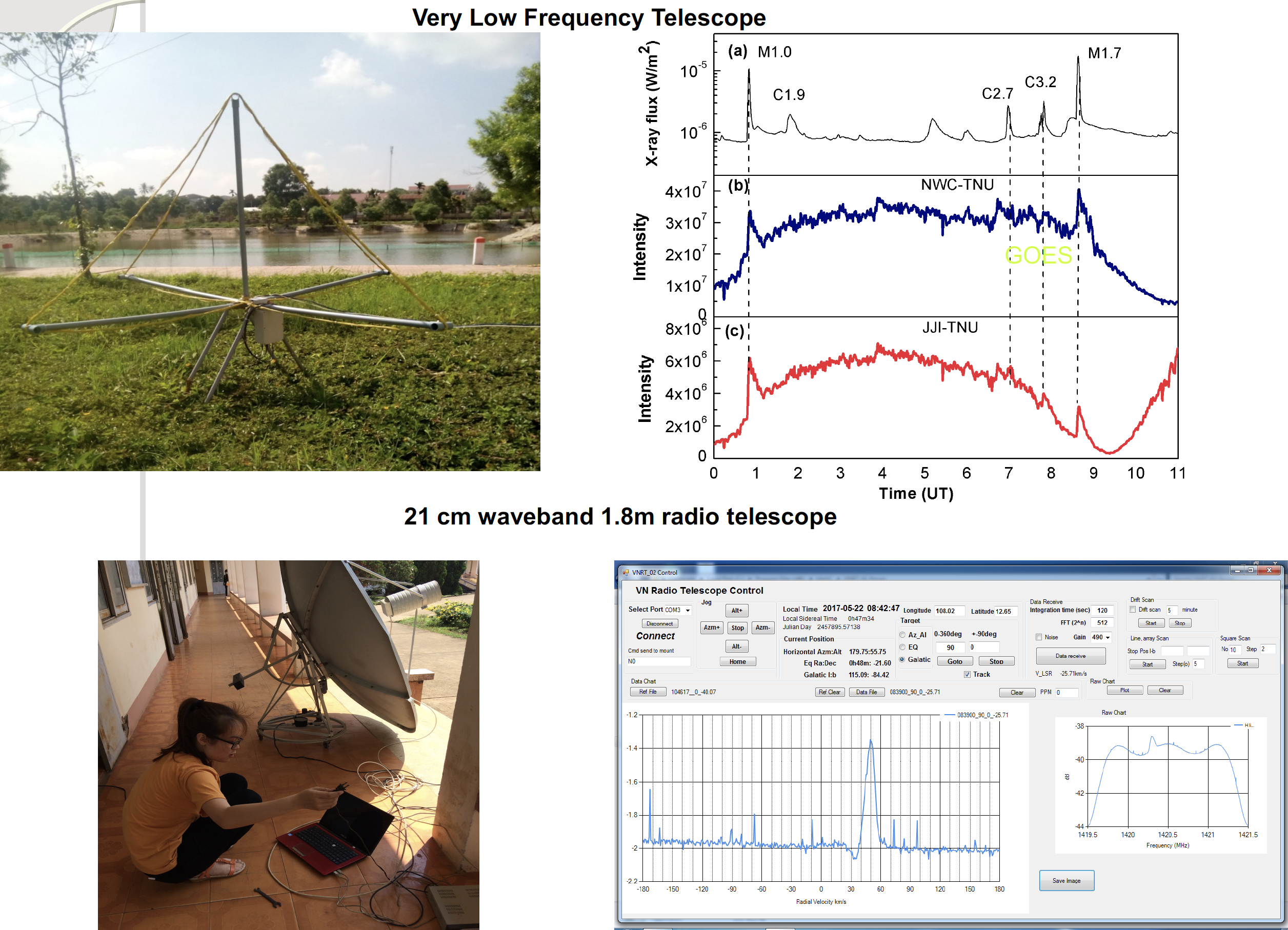}}
\caption[]{Current astronomical instruments at TNU: Very Low Frequency Array (upper) and 21 cm waveband radio telescope (lower).}
\label{fig:radish}
\end{figure}

\begin{itemize}

\item	Construct an astronomical observatory for education by 2023

\item	Build an astronomy hub at TNU for local researchers to meet and exchange continuously

\item	Nurture young enthusiasts and encourage students to follow higher degree in astronomy

\end{itemize}

\section{Main current initiative and relevant activities}
\subsection{The main activities}
This project will be a long term (3-5 years) project and will be supported by institutes such as National Astronomical Observatory of Japan (NAOJ), Kyoto Univeristy, Nishi-Harima Astronomical Observatory (NHAO), etc. The main activities include three parts including exchange staffs, on-site training, and building construction. 

{\bf Exchange staffs:} There are two main parts within this activity. First, we will send TNU staffs to Kyoto university to conduct experimentation and construction with Kyoto university staff, the cost is covered jointly by TNU and Kyoto university. Second, Japanese researchers will visit TNU to give lecture, exchange, promote students. This activity is supported mainly by TNU fund and NAOJ fund. 

{\bf On-site training:} We are planning for a 1-week school in telescope training on-site at TNU in 2019 after part of the telescope is finished. 

{\bf Observatory construction:} A building that includes housing for the telescope, control room, museum room is already planned and will be the responsibility of TNU. The TNU has agreed to support this activity. Kyoto university will support with the telescope tube and the overall technical design of the telescope. The observatory will also host the current existing research and training instruments at TNU such as Very Low Frequency telescope and 21 cm waveband radio telescope. We have used these instruments effectively in research and training in the past years$^{[4,5,6,7,8,9]}$.

\subsection{Past relevant activities}

The following past activities promoted the education and research activities in Astronomy at TNU,

\begin{itemize}
\item First visit of Nguyen Luong Quang (CITA) in 2015 to TNU has promoted students in astronomy studying 

\item In December 2015, Le Minh Tan (TNU) visited to Nobeyama Radio Observatory (NAOJ), Japan for observational practice.

\item In July 2016, Professor Masatoshi Ohishi (NAOJ) and Nguyen Luong Quang gave lectures about a basic of radio astronomy and promoted the observatory construction project at TNU.

\item In March 2017, Nguyen Dang Thanh Nhan (TNU) visited NHAO, University of Hyogo, Japan to learn how to use the 60 cm optical telescope to conduct telescopes, observations of asteroids, and analyze the data.

\item In July 2017, Professor Neal Evans (The University of Texas at Austin) visited TNU and gave students the basics of Astrophysics.

\item In June 2017, Le Minh Tan (TNU) visited Mitaka (NAOJ), Japan to learn the ALMA data analysis.

\item Tran Quoc Lam (TNU) joined the NARIT Optical Design Summer School 2018 in Chiang Mai, Thai Land

\end{itemize}

\subsection{Educational activities}
\begin{itemize}
\item	Tropical school for undergraduate astronomy at ICISE, Qui Nhon, Vietnam in April 2017
\end{itemize}

We were the first university to partner with ICISE to bring students to ICISE and organize the first hands-on event in ICISE in April, 2017. We are looking to strengthen this partnership and creating a better model for university and ICISE collaboration. We created a tropical school series in astronomy with lectures from ICISE conference attendants. School had 60 attendants (see Figure 2). We also organized a night-sky observation in ICISE's garden using a 40 cm telescope and some small telescopes designed by Le Quang Thuy and Nguyen Anh Tuan (ExploraSciene Center, ICISE) (see Figure 3). Students were very interested in this activity and really impressed with seeing the bright stars through the optical telescope.

\begin{figure}[!htbp]
\centerline{\includegraphics[width=0.9\linewidth]{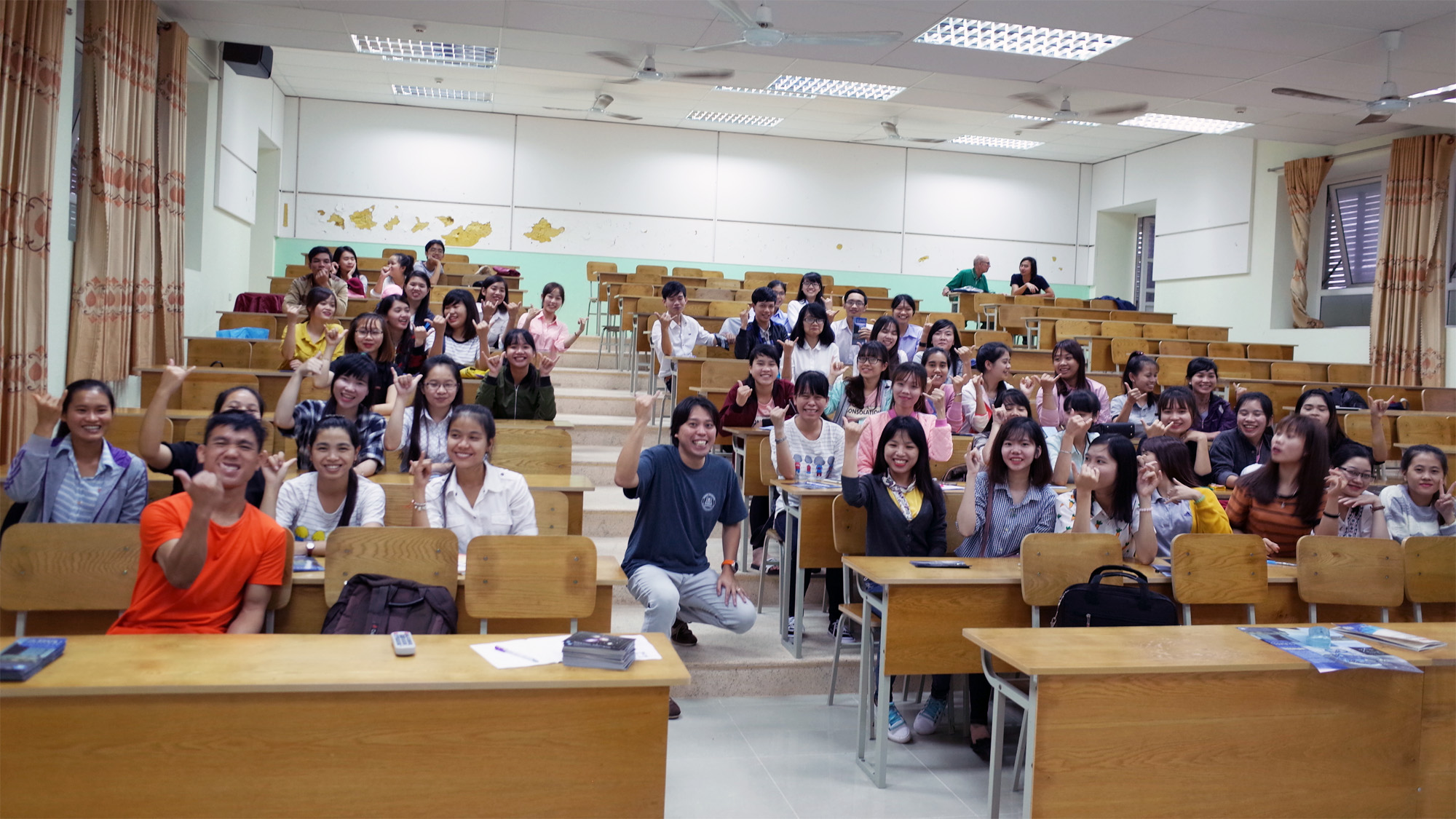}}
\caption[]{Dr. Hideaki Fujiwara (NAOJ) and TNU students in the tropical school for undergraduate astronomy in Qui Nhon university}
\label{fig:radish}
\end{figure}

\begin{figure}[htbp]
\centerline{\includegraphics[width=0.9\linewidth]{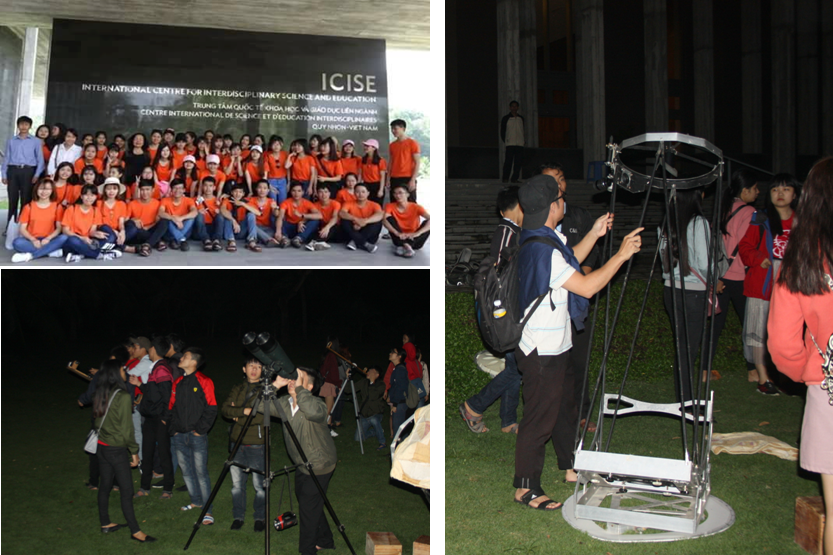}}
\caption[]{Students attended the sky observation section at ICISE garden}
\label{fig:radish}
\end{figure}

\begin{itemize}
\item	Exploration Program in Astrophysics Research (EPAR) at TNU, 2 - 13 July, 2018
\end{itemize}

\begin{figure}[htbp]
\centerline{\includegraphics[width=1\linewidth]{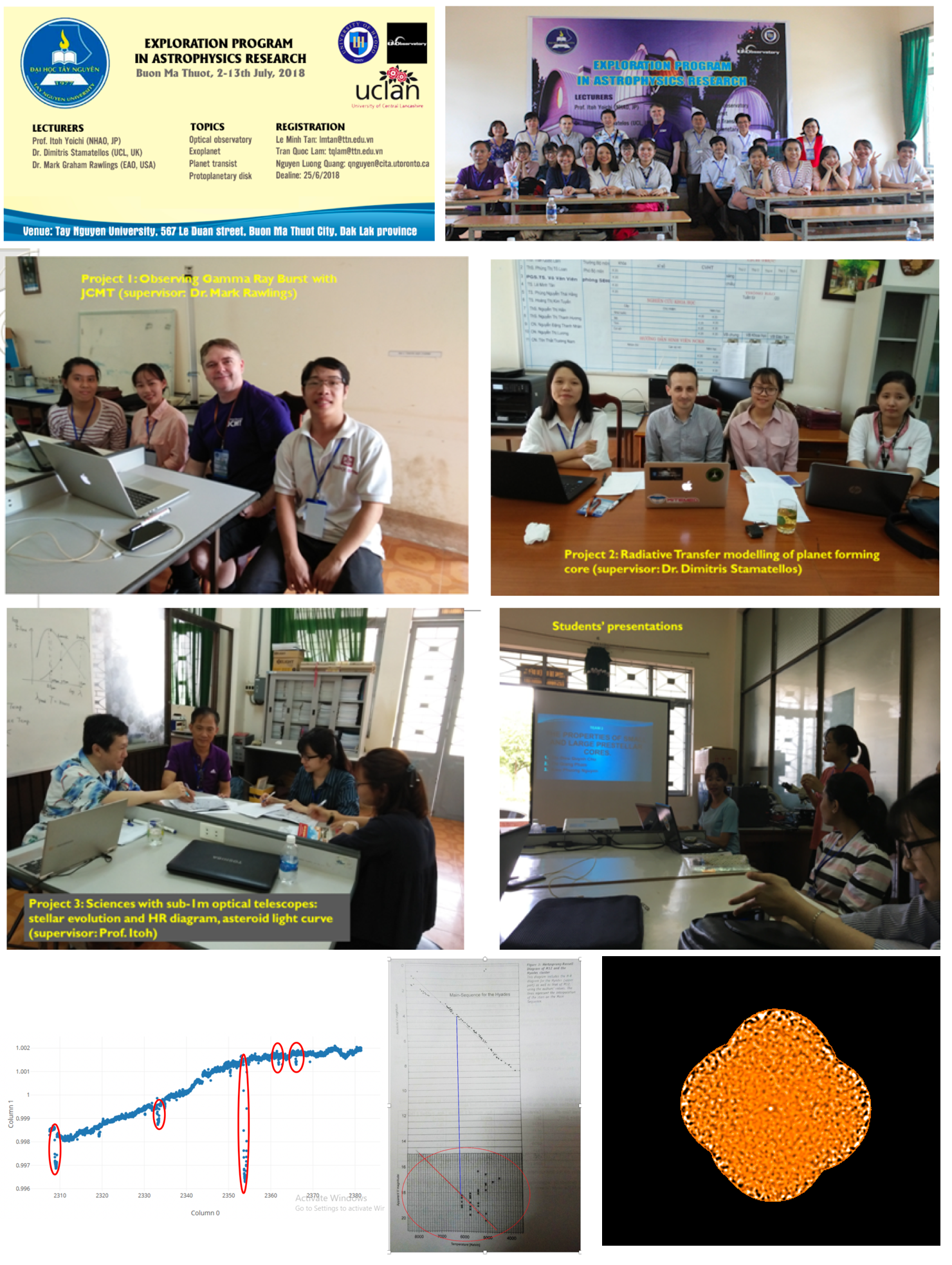}}
\caption[]{Summary of EPAR activities. Logo of EPAR and opening section of EPAR (upper), student groups were working on their projects (middle lowers), and results from individual working group (lower): exoplanet transist (left), HR diagram (middle), and gamma-ray burst (right).}
\label{fig:radish}
\end{figure}

The EPAR at TNU brings astronomers and students together to work on specific research topics in astrophysics. The program run for 2 weeks, hosted by TNU. The program welcomed students from all parts of Vietnam who are senior undergraduate students or graduate students searching for research experiences. Vietnamese students who are searching for a place to host their dissertations are also welcomed. Participants can come for a shorter period or for the whole duration of the program, and are encouraged to come back year after year. Three visiting scientists and nine students teamed up and conduct common research projects (see Figure 5). The projects were designed such that they can be sub-divided into individual sub-projects and tailored to match with the availability of the supervisor that year. Each week, scientists were responsible for the scientific theme and student supervision. Students presented their progress in the two Friday's afternoons. These projects are described with the details below,

\textit{Project 1: Observing Gamma Ray Burst with James Clerk Maxwell Telescope (JCMT) (supervisor: Dr. Mark Rawlings, EAO, USA)}

Students get to act as members of a GRB monitoring team. The project walks the students through the notification process for a GRB trigger, encourages them to check the position with Set of Identifications, Measurements, and Bibliography for Astronomical Data (SIMBAD), prepare a simple JCMT MSB with the JCMT Observing Tool, receive the data, examine it, conduct flux calibration and plot a simple light curve.

\textit{Project 2: Radiative Transfer modeling of planet forming core (supervisor: Dr. Dimitris Stamatellos, UCL, UK)}

In this project, students find the temperature of the prestellar cores, do radiative transfer simulations and test the cores by comparing it with previous works. Students also study the motion of the exoplanets observed by the Kepler space telescope using the web browser and directly download the data of the planet to study. Students look at some of the exoplanets of the K2 mission and determine what kind of planets they are (hot planet, cold planet or habitable planet) by defining parameters of planets such as radius, temperature, mass,....Students also calculate the parameters of orbital radius, orbital period, stellar luminosity.

\textit{Project 3: Sciences with sub-1m optical telescopes: stellar evolution and HR diagram, asteroid light curve (supervisor: Prof. Itoh Yoichi, NHAO, JP)}

Real photometric data taken with the 2 m Nayuta Telescope, a 60 cm Nishimura telescope, and a 30 cm Meade telescope at NHAO will be examined. The exercise consists of three steps; photometry of open cluster members, time series photometry of an asteroid, and time series precise photometry of an exoplanet transit.

\begin{itemize}
\item	Observational event at Nha Trang Observatory (NTO) from 4 - 7 November 2018
\end{itemize}

\begin{figure}[h!]
\centerline{\includegraphics[width=0.87\linewidth]{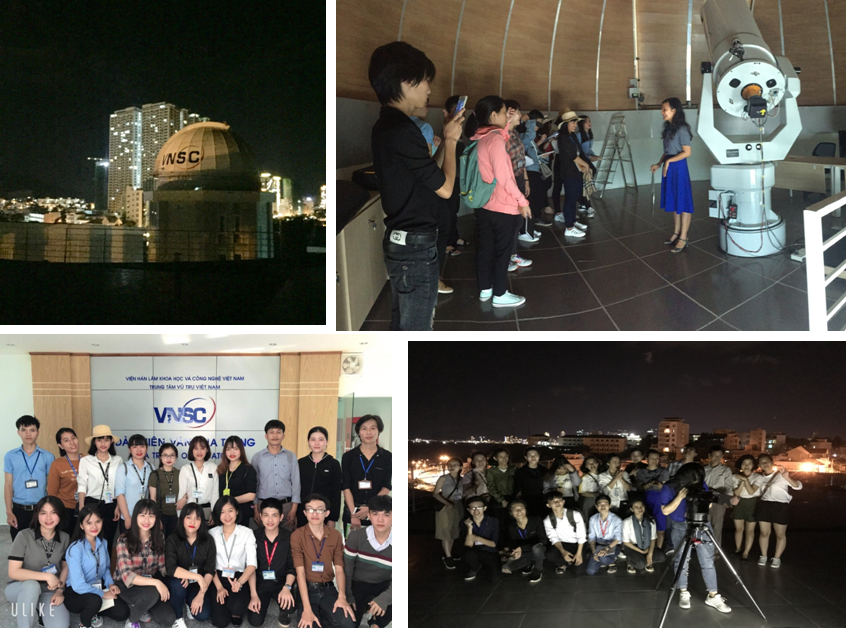}}
\caption[]{Student activities in observational event at NTO}
\label{fig:radish}
\end{figure}

Figure 6 shows the activities of TNU students at NTO. Students surveyed the parts of the 50 cm optical telescope to understand how it work and were provided with the basic knowledge of the sky. The lectures presented by staff of NTO focused on the basic concepts and guidelines for observing the sky. Students did the exercises on how to determine the zodiac constellations, and identify the constellations in the summer triangle and typically bright stars in particular season. Students also experienced and admired the beauty of the universe at the planetarium with film ``Two small pieces of glass". In the evening students experienced the night sky with a small optical telescope to practice the knowledge which they were learnt in the planetarium.

\section{Future plan}

\subsection{Strategic plans}
\begin{itemize}
\item	Establish annual EPAR program at TNU and in corporation with other institutes
\item Exploit scientific projects that can be done with 40cm telescopes
\item Incorporate the hands-on activities with telescopes into astronomy curriculum
\item Involve in extended training and research network
\end{itemize}

\subsection{Detailed plans}
\begin{itemize}
\item {\bf  2019: }
We will continue to organize the EPAR at TNU and NTO, in July 2019. It will run for two weeks. This program will focus on the topics such as basics of Astrophysics, photometry and spectroscopy. In the second week, the school will conduct astrophysical observation training for university students and young scientists using the 50cm optical telescope at NTO. The number of participants is about 15. The observation targets are the dispersed star and asteroids.
At the end of 2019, we will start to build the observatory. The site area is about 100 m$^{2}$ inside TNU campus.

\item {\bf  2020-2021: }
Dr. Mikio Kurita, an associate professor at Kyoto University, will establish a 40-cm telescope at TNU. This telescope will be installed, operating, and so on, to install CCD cameras and to conduct initial testing observations. 
We will organize the Astronomy training. Students will conduct observation on university students from the previous fiscal year (2019) using the 50cm optical telescope at NTO. This training was conducted using newly developed optical spectrometer.

\item {\bf 2022-2023: }
Currently, VNSC has two observatories. We hope that we will join the optical telescope network in Vietnam. In addition, the TNU 40 cm telescope will be also included in an observation network of Japan to participate in the observation of unexpected celestial bodies such as supernova explosions and gravitational waves.
\end{itemize}

\section{Summary and Challenges}

We plan to build an astronomical observatory at TNU and to design a 40cm telescope with the new design led by the astronomy lab from Kyoto university. Land and observatory construction cost is granted by TNU.
We successfully organized ``Tropical school for undergraduate astronomy" at ICISE in Qui Nhon and ``Observational event" in NTO. These activities aim to provide students with the skills of night sky observation and to create their passion in astronomy study. 
We successfully organized the EPAR from July 2 to July 13, 2018 at TNU. The goals of this program are to promote the teaching and learning activities on Astrophysics, and to promote the astronomical observatory construction project at TNU.
We also showed the future plan for observatory construction. 

However, we meet many challenges such as human source, budget and collaboration. There is a lack of experts in Astronomy in Central Highland, Vietnam. We are actively looking for financial support to purchase parts of the observatory and send staffs to visit foreign institutes. We actively need support from scientific community to come to Daklak to teach and exchange. We are also looking for support to receive our students for internship or exchange. 

\vskip 0.5cm

{\bf Acknowledgement: }
During the course of our work, we appreciate the supports from individuals such as tropical school lecturers, EPAR lecturers, Kaz Sekiguchi, Mikio Kurita, Yoichi Itoh, Neal Evans, Yong-Chol Minh, Hyung Mok Lee, Nguyen Thi Hoang Anh, Pham Ngoc Diep, Nguyen Trong Hien and institutes such as NAOJ, ICISE, KASI, NHAO, Kyoto University, VNSC, IAU-OAD.

\end{document}